\documentclass[12pt]{article}
\usepackage{amsfonts}

\usepackage{amsmath}
\usepackage{cite}
\usepackage{graphicx}
\csname @addtoreset\endcsname{equation}{section}
\newcommand{\eq}{\begin{equation}}
\newcommand{\feq}{\end{equation}}
\newcommand{\eqn}{\begin{eqnarray}}
\newcommand{\feqn}{\end{eqnarray}}
\newcommand{\arr}{\begin{eqnarray*}}
\newcommand{\farr}{\end{eqnarray*}}

\font\mybb=msbm10 at 12pt
\def\bb#1{\hbox{\mybb#1}}

\def\bR {\bb{R}}

\def\bC {\bb{C}}

\begin{document}

\begin{titlepage}
\begin{flushright}
IFUM-694-FT\\
hep-th/0110031
\end{flushright}
\vspace{.3cm}
\begin{center}
\renewcommand{\thefootnote}{\fnsymbol{footnote}}
{\Large \bf The Asymptotic Dynamics of de~Sitter Gravity in three Dimensions}
\vfill
{\large \bf {Sergio Cacciatori\footnote{cacciatori@mi.infn.it} and
Dietmar Klemm\footnote{dietmar.klemm@mi.infn.it}}}\\
\renewcommand{\thefootnote}{\arabic{footnote}}
\setcounter{footnote}{0}
\vfill
{\small
Dipartimento di Fisica dell'Universit\`a di Milano
and\\ INFN, Sezione di Milano,
Via Celoria 16,
20133 Milano, Italy.\\}
\end{center}
\vfill
\begin{center}
{\bf Abstract}
\end{center}
{\small We show that the asymptotic dynamics of three-dimensional gravity
with positive cosmological constant is described by Euclidean Liouville
theory. This provides an explicit example of a correspondence
between de~Sitter gravity and conformal field theories.
In the case at hand, this correspondence is established by formulating
Einstein gravity with positive cosmological constant in three dimensions as
an SL$(2,\bC)$ Chern-Simons theory. The de~Sitter boundary conditions
on the connection are divided into two parts. The first part reduces
the CS action to a nonchiral SL$(2,\bC)$ WZNW model,
whereas the second provides the constraints for a further reduction
to Liouville theory, which lives on the past boundary of dS$_3$.}

\end{titlepage}

\section{Introduction}

Recently there has been an increasing interest in gravity
on de~Sitter (dS) spacetimes \cite{Hull:1998vg,Hull:2000mt,Spradlin:2001pw,
Banks:2000fe,Balasubramanian:2001rb,Witten:2001kn,Bousso:2000nf,
Volovich:2001rt,
Chamblin:2001dx,Strominger:2001pn,Klemm:2001ea,Mazur:2001aa,
Li:2001ky,Nojiri:2001mf,Gao:2001sr,Shiromizu:2001bg,Kallosh:2001tm,
Hull:2001ii}.
The motivation for this comes partially from recent astrophysical
data indicating a positive cosmological constant \cite{Perlmutter:2000rr},
but apart from phenomenological aspects,
it also remains an outstanding challenge to understand the role
of de~Sitter space in string theory, to clarify the microscopic
origin of de~Sitter entropy \cite{Maldacena:1998ih,Park:1998qk,Banados:1999tb,
Kim:1999zs,Lin:1999gf,Hawking:2001da},
and to study in which way the holographic
principle \cite{'tHooft:1993gx,Susskind:1995vu} is realized in the
case of de~Sitter gravity. Whereas
string theory on anti-de~Sitter spaces is known to have a dual description
in terms of certain superconformal field theories \cite{Maldacena:1998re},
no such explicit duality was known up to now for dS spacetimes.
Based on \cite{Hull:1998vg}, where the first evidence for a dS/CFT
correspondence was given, and on related ideas that appeared
in \cite{Hull:2000mt,Bousso:1999cb,Balasubramanian:2001rb},
Strominger proposed recently a holographic duality
relating quantum gravity on dS$_D$ to a Euclidean conformal
field theory residing
on the past boundary ${\cal I}^-$ of dS$_D$ \cite{Strominger:2001pn}.
In particular, he considered three-dimensional de~Sitter gravity,
and showed that it admits an asymptotic symmetry algebra consisting
of two copies of Virasoro algebras with central charges
$c = \tilde c = 3l/2G$, where $l$ is the dS$_3$ curvature radius, and
$G$ denotes Newton's constant\footnote{This generalizes the result of
Brown and Henneaux \cite{Brown:1986nw} to the case of positive
cosmological constant.}. This central charge was then rederived by
other methods in \cite{Klemm:2001ea}, where also further evidence for a dS/CFT
correspondence was given.\\
The purpose of the present paper is to show that in the case of
$2+1$-dimensional Einstein gravity
with positive cosmological constant, the dual CFT in question is
Euclidean Liouville field theory\footnote{Cf.~also \cite{Deser:dr,Mazur:qc}
for the appearance of the Liouville equation and Liouville action in
slightly different contexts.}.
This provides an explicit example of
a dS/CFT correspondence, and shows how the asymptotic symmetries of
dS$_3$ are realized in the dual conformal field theory.
To establish this correspondence, we shall make use of the Chern-Simons
formulation \cite{Achucarro:1986vz,Witten:1988hc}
of pure gravity in $2+1$ dimensions, and
we will closely follow the analogous work that has been
done for anti-de~Sitter space \cite{Coussaert:1995zp,Banados:1998pi,
Henneaux:2000ib,Rooman:2001zi,Krasnov:2001cu}.\\
The remainder of this paper is organized as follows:\\
In the next section, we briefly review the Chern-Simons formulation
of three-dimensional Einstein gravity with positive cosmological constant,
and translate the de~Sitter boundary conditions on the
metric \cite{Strominger:2001pn} in terms of the CS connection.
In section \ref{redWZNW}, we show that the first part of these conditions
implies the reduction of the Chern-Simons action to a nonchiral
SL$(2,\bC)$ WZNW model.
In \ref{redLiou}, we make use of the remaining boundary conditions
in order to further reduce the WZNW model to Liouville field theory,
which lives on the past boundary ${\cal I}^-$ of dS$_3$.
The appendix contains our conventions.

\section{De~Sitter Gravity as Chern-Simons Theory}
\label{dSCS}

Pure $2+1$-dimensional gravity with positive cosmological constant
$\Lambda = 1/l^2$ is described by the action

\begin{equation}
S = \frac{1}{16\pi G} \int d^3x \sqrt{-g}(R - 2\Lambda)\,. \label{gravaction}
\end{equation}

The equations of motion following from (\ref{gravaction}) admit the
de~Sitter solution\footnote{This is related to the planar slicing
of dS$_3$ given in \cite{Strominger:2001pn} by setting $\tau = l\exp(-t)$.
The past boundary ${\cal I}^-$ ($t \to -\infty$) corresponds then to
$\tau \to \infty$. The Carter-Penrose diagram can be found
in \cite{Strominger:2001pn}.}

\begin{equation}
ds^2 = -\frac{l^2}{\tau^2}d\tau^2 + \tau^2 dzd\bar{z}\,. \label{dSsol}
\end{equation}

An asymptotically de~Sitter geometry \cite{Strominger:2001pn} is
one for which the metric behaves for $\tau \to \infty$ as

\begin{eqnarray}
g_{z\bar{z}} &=& \frac{\tau^2}{2} + {\cal O}(1)\,, \nonumber \\
g_{zz} &=& {\cal O}(1)\,, \nonumber \\
g_{\tau\tau} &=& -\frac{l^2}{\tau^2} + {\cal O}(\tau^{-4})\,,
                 \label{boundcondmetr} \\
g_{z\tau} &=& {\cal O}(\tau^{-3})\,. \nonumber
\end{eqnarray}

Past infinity ${\cal I}^-$ is a spacelike cylinder
parametrized by the coordinates $\phi \sim \phi + 2\pi$ and $y$,
where we set $z = \phi + iy$. We denote this surface by $\Sigma$.\\

In what follows, we shall make essential use of the fact that
$2+1$-dimensional de~Sitter gravity can be formulated as an
SL$(2,\bC)$ Chern-Simons
theory \cite{Achucarro:1986vz,Witten:1988hc}, with action

\begin{equation}
S = \frac{is}{4\pi}\int{\mathrm{Tr}}(A\wedge dA + \frac 23
    A\wedge A\wedge A) -
    \frac{is}{4\pi}\int{\mathrm{Tr}}(\bar{A}\wedge d\bar{A} + \frac 23
    \bar{A}\wedge \bar{A}\wedge \bar{A})\,, \label{CSaction}
\end{equation}

where, in our conventions (cf.~appendix),

\begin{equation}
s = -\frac{l}{4G}\,,
\end{equation}

and $A$ denotes an SL$(2,\bC)$ gauge field. (\ref{gravaction}) and
(\ref{CSaction}) are equivalent, if we identify

\begin{equation}
A = A^a\tau_a = \left(\omega^a + \frac il e^a\right)\tau_a\,, \qquad
\bar A = {\bar A}^a\tau_a = \left(\omega^a - \frac il e^a\right)\tau_a\,,
\quad a = 0,1,2\,, \label{defA}
\end{equation}

where $e^a$ is the dreibein,
$\omega^a = \frac 12 \epsilon^{abc}\omega_{bc}$
the SL$(2,\bR)$ spin connection, and the $\tau_a$ are the SL$(2,\bC)$
generators (cf.~appendix).\\
Modulo total derivatives\footnote{Boundary terms will be considered below.},
the action (\ref{CSaction}) can be rewritten as

\begin{equation}
S = S_{CS}[A] - S_{CS}[\bar A]\,, \label{newCSaction}
\end{equation}

with

\begin{equation}
S_{CS}[A] = \frac{is}{4\pi}\int d^3x\, {\mathrm{Tr}}\,(\dot A_{\tau} A_{\phi}
            - \dot{A}_{\phi} A_{\tau} + 2 A_y F_{\tau\phi})\,,
\end{equation}

where a dot denotes the derivative with respect to $y$, and $F = dA +
A\wedge A$.\\
In terms of the connection $A$, the boundary conditions
(\ref{boundcondmetr}) read

\begin{equation}
A = \left(\begin{array}{cc} \frac{d\tau}{2\tau} & {\cal O}(1/\tau) \\
                            & \\
                            \frac{i\tau}{l}dz & -\frac{d\tau}{2\tau}
                            \end{array}\right)\,, \qquad
\bar A = \left(\begin{array}{cc} -\frac{d\tau}{2\tau} & -\frac{i\tau}{l}
                                 d\bar z \\
                                 & \\
                                 {\cal O}(1/\tau) & \frac{d\tau}{2\tau}
                            \end{array}\right)\,. \label{asbehav}
\end{equation}

Like in the AdS case \cite{Coussaert:1995zp}, we can state two essential
points concerning the asymptotic behaviour (\ref{asbehav}) of the
connection:
\begin{enumerate}
\item The components $A_{\bar z}$ and ${\bar A}_z$ are set to zero
asymptotically.
\item $A^+_z = A^1_z + iA^2_z$ and ${\bar A}^-_{\bar z} = {\bar A}^1_{\bar z}
- i{\bar A}^2_{\bar z}$ are independent of $z, \bar z$ to leading order
in $\tau$. Also, $A^0_z$ and ${\bar A}^0_{\bar z}$ vanish.
\end{enumerate}

In section \ref{redWZNW}, following \cite{Coussaert:1995zp,Moore:1989yh},
we show that the first condition implies
a reduction of the Chern-Simons action to a nonchiral WZNW model.
The second condition then provides the constraints for a further
reduction from the WZNW model to Liouville theory.

\section{Reduction to a WZNW Model}
\label{redWZNW}

When $A_{\bar z}$ and ${\bar A}_z$ are required to vanish on the boundary
${\cal I}^- = \Sigma$, the action (\ref{newCSaction}) is not extremal
on-shell. Instead, $\delta S$ equals the surface term
$\delta [-\frac{s}{4\pi}\int_{\Sigma}d\phi dy\, {\mathrm{Tr}}(A_{\phi}^2
+ {\bar A}_{\phi}^2)]$. In order to cancel this, we must add a boundary
term to the action (\ref{newCSaction}), leading to the improved
action\footnote{Surface terms that arise at $y_1$ and $y_2$ will be discussed
in section \ref{redLiou}. Besides, de~Sitter space has two spacelike
conformal boundaries ${\cal I}^-$ and ${\cal I}^+$, so that in principle
additional boundary terms at ${\cal I}^+$ have to be taken into account.
However, since we are only interested in the asymptotic dynamics
of the gravitational field near ${\cal I}^-$, we will ignore such
surface terms.
Some discussion of the problems that arise due to the presence
of a second boundary
can be found in \cite{Strominger:2001pn},
where it was argued that in spite of the fact that de~Sitter space has
two boundaries, the dual description of de~Sitter gravity is provided
by a single CFT.}

\begin{equation}
S = S_{CS}[A] + \frac{s}{4\pi}\int_{\Sigma}d\phi dy\, {\mathrm{Tr}}\,
    [(A_{\phi})^2]
    -S_{CS}[\bar A] + \frac{s}{4\pi}\int_{\Sigma}d\phi dy\,
    {\mathrm{Tr}}\,[({\bar A}_{\phi})^2]\,. \label{improved}
\end{equation}

We recognize that $A_y$ and ${\bar A}_y$ are Lagrange multipliers that
implement the Gauss law constraints $F_{\tau\phi} = {\bar F}_{\tau\phi} = 0$.
These are easily solved by\footnote{We did not consider possible
holonomies, that appear as additional zero modes in (\ref{solvconstr}).}

\begin{equation}
A_{\mu} = G_1^{-1}\partial_{\mu}G_1\,, \qquad
{\bar A}_{\mu} = G_2^{-1}\partial_{\mu}G_2\,, \label{solvconstr}
\end{equation}

where $G_{1,2}$ have the asymptotic behaviour

\begin{equation}
G_1 \to g_1(z, \bar z) \left(\begin{array}{cc} \sqrt{\frac{\tau}{l}} & 0 \\
                                                            &   \\
                                            0 & \sqrt{\frac{l}{\tau}}
                          \end{array} \right)\,, \qquad
G_2 \to g_2(z, \bar z) \left(\begin{array}{cc} \sqrt{\frac{l}{\tau}} & 0 \\
                                                            &   \\
                                            0 & \sqrt{\frac{\tau}{l}}
                          \end{array} \right)\,, \label{asG}
\end{equation}

and $g_{1,2}(z, \bar z)$ are arbitrary elements of SL$(2,\bC)$.
Strictly speaking, the Gauss law constraints imply (\ref{solvconstr})
only for $\mu = \tau,\phi$, whereas (\ref{solvconstr}) for $\mu = y$
is a gauge choice.
(\ref{asG}) implies the boundary condition (\ref{asbehav}) for
$A_{\tau}$ and ${\bar A}_{\tau}$, whereas for the tangential components
one obtains

\begin{equation}
A_j = \left(\begin{array}{cc} -\frac i2 a^0_j & \frac{l}{2\tau} a^-_j \\
                                              &                       \\
                              \frac{\tau}{2l} a^+_j & \frac i2 a^0_j
                              \end{array} \right)\,, \qquad
{\bar A}_i = \left(\begin{array}{cc} -\frac i2 {\tilde a}^0_j &
                    \frac{\tau}{2l} {\tilde a}^-_j \\ & \\
                    \frac{l}{2\tau} {\tilde a}^+_j & \frac i2 {\tilde a}^0_j
                              \end{array} \right)\,,
\end{equation}

where $a_j \equiv g_1^{-1}\partial_j g_1$, ${\tilde a}_j \equiv
g_2^{-1}\partial_j g_2$, and $j = z,\bar z$.\\
Inserting (\ref{solvconstr}) into the action (\ref{improved})
yields a sum of two chiral WZNW models,

\begin{equation}
S = S^R_{WZNW}[g_1] - S^L_{WZNW}[g_2]\,, \label{sum}
\end{equation}

where

\begin{eqnarray}
S^R_{WZNW}[g_1] &=& \frac{s}{2\pi} \int d\phi dy\,{\mathrm{Tr}}\,[g_1^{-1}
                    \partial_{\bar z}g_1 g_1^{-1}\partial_{\phi}g_1] -
                    \frac{is}{12\pi}\int {\mathrm{Tr}}\,(G_1^{-1}dG_1)^3\,,
                    \nonumber \\
S^L_{WZNW}[g_2] &=& -\frac{s}{2\pi} \int d\phi dy\,{\mathrm{Tr}}\,[g_2^{-1}
                    \partial_z g_2 g_2^{-1}\partial_{\phi}g_2] -
                    \frac{is}{12\pi}\int {\mathrm{Tr}}\,(G_2^{-1}dG_2)^3\,.
\end{eqnarray}

These first order actions describe respectively a holomorphic group
element $g_1(z)$ and an antiholomorphic group element $g_2({\bar z})$.
One has thus $a_{\bar z} = {\tilde a}_z = 0$ on-shell,
so that the first part of
the boundary conditions is indeed satisfied.\\
The sum (\ref{sum}) of right- and left chiral actions is equivalent
to the nonchiral WZNW action with dynamical variable $g = g_1^{-1}g_2$.
To see this equivalence, we rewrite the action (\ref{sum}) in terms
of the new variables $g$ and $\Pi \equiv -g_2^{-1}\partial_{\phi}g_1
g_1^{-1}g_2 - g_2^{-1}\partial_{\phi}g_2$. This leads to

\begin{equation}
S = \frac{s}{2\pi}\int d\phi dy\, {\mathrm{Tr}}\left[\frac i2 \Pi g^{-1}
    \partial_y g + \frac 14 \Pi^2 + \frac 14 g^{-1}\partial_{\phi}g
    g^{-1}\partial_{\phi}g\right] + \frac{is}{12\pi}\int {\mathrm{Tr}}\,
    (G^{-1}dG)^3\,, \label{WZNW1storder}
\end{equation}

with $G = G_1^{-1}G_2$. (\ref{WZNW1storder}) is exactly the nonchiral
WZNW model in first order formalism. Eliminating the auxiliary
variable $\Pi$ by its equation of motion, one gets finally

\begin{equation}
S = \frac{s}{2\pi}\int d\phi dy\, {\mathrm{Tr}}\,[g^{-1}\partial_z g
    g^{-1}\partial_{\bar z} g] + \frac{is}{12\pi}\int {\mathrm{Tr}}\,
    (G^{-1}dG)^3\,, \label{WZNW}
\end{equation}

which is the standard WZNW action.

\section{Further Reduction to Liouville Theory}
\label{redLiou}

Up to now, we have implemented only part 1 of the boundary conditions
on the Chern-Simons connection. We must still incorporate the second
part, which, in terms of the Kac-Moody currents, read

\begin{equation}
J_{\bar z}^- \equiv (g^{-1}\partial_{\bar z} g)^- = -2i\,, \qquad
\tilde{J}_z^+ \equiv (\partial_z g g^{-1})^+ = -2i\,, \label{constraints}
\end{equation}

and

\begin{equation}
J_{\bar z}^0 = \tilde{J}_z^0 = 0\,. \label{gaugecond}
\end{equation}

The constraints (\ref{constraints}) are first class among themselves,
and therefore generate a gauge symmetry, while the conditions
(\ref{gaugecond}) can be viewed as gauge conditions for the symmetry
generated by (\ref{constraints}) \cite{Henneaux:2000ib}.
If one implements (\ref{constraints}) and (\ref{gaugecond}) by means of
Lagrange multipliers, one gets the (gauged-fixed version of the) action
for the gauged WZNW model, in which one has gauged the subgroup
of SL$(2,\bC)$ generated
by the first class constraints \cite{Henneaux:2000ib}. It is well
known \cite{Alekseev:1989ce,Bershadsky:1989mf,Forgacs:1989ac,Sabra:1993hw}
that this model is equivalent to Liouville theory.
To see this equivalence at the level of the action, we parametrize
$g \in$ SL$(2,\bC)$ according to the Gauss decomposition

\begin{equation}
g = \left(\begin{array}{cc} 1 & X \\ & \\ 0 & 1 \end{array}\right)
    \left(\begin{array}{cc} \exp(\frac 12 \Phi) & 0 \\ & \\
     0 & \exp(-\frac 12 \Phi) \end{array}\right)
    \left(\begin{array}{cc} 1 & 0 \\ & \\ Y & 1 \end{array}\right)\,,
    \label{decomp}
\end{equation}

where $X, Y, \Phi \in \bC$. It will be shown below that actually $\Phi$
is real. With (\ref{decomp}), the action (\ref{WZNW})
reads \cite{Forgacs:1989ac}

\begin{equation}
S = \frac{s}{2\pi}\int d\phi dy \left[\frac 12 \partial_z \Phi
    \partial_{\bar z} \Phi + 2 \partial_{\bar z} X \partial_z Y \exp(-\Phi)
    \right]\,. \label{preLiouville}
\end{equation}

In terms of the "momenta" $\Pi_X = \partial {\cal L}/\partial \dot X$
and $\Pi_Y = \partial {\cal L}/\partial \dot Y$ conjugate to $X, Y$, one
obtains for the constraints (\ref{constraints})

\begin{equation}
\Pi_X = \frac{is}{2\pi}\partial_z Y\exp(-\Phi) = \frac{s}{2\pi}\,, \qquad
\Pi_Y = -\frac{is}{2\pi}\partial_{\bar z} X\exp(-\Phi) = -\frac{s}{2\pi}\,.
        \label{momconstr}
\end{equation}

In order to implement (\ref{momconstr}), we have to go from
(\ref{preLiouville}) to the reduced action (Routhian function),

\begin{equation}
S \to S - \int d\phi dy \left[\dot X \Pi_X + \dot Y \Pi_Y\right]\,,
      \label{redact}
\end{equation}

i.~e.~we have to perform a partial Legendre transformation with respect to
$\dot X, \dot Y$. This is equivalent to the procedure used in
\cite{Coussaert:1995zp}, which consists in adding a boundary term to
(\ref{preLiouville}), to get an improved action

\begin{equation}
S_{impr} = S + \frac{is}{2\pi}\oint d\phi \left.\left[Y \partial_{\bar z} X
           - X \partial_z Y\right]\exp(-\Phi)\right|^{t_2}_{t_1}\,.
           \label{impract}
\end{equation}

One can then insert (\ref{momconstr}) into (\ref{redact}) or
(\ref{impract}), to obtain finally

\begin{equation}
S = \frac{s}{2\pi} \int d\phi dy \left[\frac 12 \partial_z \Phi
    \partial_{\bar z} \Phi + 2\exp\Phi\right]\,, \label{Liouville}
\end{equation}

which is the action of Euclidean Liouville field theory.
We have thus shown that the asymptotic dynamics of three-dimensional
de~Sitter gravity is described by Liouville field theory. This provides
an explicit example of the dS/CFT correspondence proposed by
Strominger \cite{Strominger:2001pn}. The two sets of Virasoro generators
of Liouville theory are related to the residual Kac-Moody symmetries
preserving the constraints (\ref{constraints}) \cite{Coussaert:1995zp,
Forgacs:1989ac}, given by

\begin{equation}
g(z, \bar z) \to \Omega(z)g(z, \bar z)\tilde{\Omega}^{-1}(\bar z)\,,
\end{equation}

with

\begin{equation}
\Omega(z) = \pm \left(\begin{array}{cc} 1 & f(z) \\ 0 & 1
            \end{array} \right)\,, \qquad
\tilde{\Omega}(\bar z) = \pm \left(\begin{array}{cc} 1 & 0 \\
            \tilde f(\bar z) & 1 \end{array} \right)\,,
\end{equation}

where $f(z)$ ($\tilde f(\bar z)$) is an arbitrary holomorphic
(antiholomorphic) function.\\

We still have to show that the Liouville mode $\Phi$ is real. This
is not evident from the Gauss decomposition (\ref{decomp}), as this
in general requires a complex $\Phi$ for the group SL$(2,\bC)$.
We did however not yet implement all the constraints.
As our SL$(2,\bC)$ generators satisfy $\tau_a^{\dag} = \sigma\tau_a \sigma$
(cf.~appendix), with $\sigma$ given in (\ref{sigma}), the identification
(\ref{defA}) leads to the pseudoreality condition

\begin{equation}
A^{\dag} = \sigma \bar A \sigma\,,
\end{equation}

which implies $G_2^{-1} = \sigma G_1^{\dag}\tau$, where $\tau$ denotes
an arbitrary constant element of SL$(2,\bC)$. If we choose
e.~g.~$\tau^{\dag} = -\tau$, we obtain the relation\footnote{The same
holds for the boundary value $g$.}

\begin{equation}
G^{\dag} = -\sigma G \sigma\,, \label{constrG}
\end{equation}

so that $G$ has the form

\begin{equation}
G = \left(\begin{array}{cc} u & w \\ -\bar w & v \end{array}\right)\,, 
    \label{Gincoset}
\end{equation}

with $u, v \in \bR$, $w \in \bC$. It is easy to see that (\ref{Gincoset})
parametrizes an element of the coset space SL$(2,\bC)$/SU$(2)$.
The Gauss decomposition of (\ref{Gincoset}) is given by (\ref{decomp})
with $Y = -\bar X$ and $\Phi$ real. We can now impose this final constraint
on the action (\ref{Liouville}) by means of a Lagrange multiplier $\lambda$,
and then integrate in the path integral over $\lambda$ and over the
imaginary part of $\Phi$, which leads to the same Liouville action, but with
real $\Phi$. Alternatively, one can implement (\ref{constrG}) already
after the first reduction step, i.~e.~, one can restrict the group
elements appearing in the WZNW action (\ref{WZNW}) to take values
in the coset space SL$(2,\bC)$/SU$(2)$. To this end, one first translates
(\ref{constrG}) into a constraint for the currents,

\begin{equation}
J_{\bar z}^{\dag} = -\sigma \tilde{J}_z \sigma\,, \label{constrcurr}
\end{equation}

which is then imposed by means of a Lagrange multiplier in the action
(\ref{WZNW}). This amounts to gauging the subgroup SU$(2)$ of SL$(2,\bC)$,
generated by the constraints (\ref{constrcurr}), i.~e.~, one obtains
the action of the SL$(2,\bC)$/SU$(2)$ gauged WZNW model.
The integration constant $\tau$ relating $G_1$ and $G_2$ determines which
subgroup is gauged. For example, the choice $\tau^{\dag} = \tau$ leads
to a gauging of the subgroup SU$(1,1)$.\\
The classical Liouville solution corresponding to de~Sitter space
in horospherical coordinates (\ref{dSsol}) is given by

\begin{equation}
\exp\Phi = \left[-uz\bar z - iwz + i\bar w\bar z + \frac{1-w\bar w}{u}
           \right]^{-2}\,, \label{Liousol}
\end{equation}

where $u \in \bR$ and $w \in \bC$ are arbitrary constants.
By a combined dilation and translation, $uz \to z + i\bar w$, (\ref{Liousol})
can be cast in the elliptic form \cite{Seiberg:1990eb}

\begin{equation}
\exp\Phi = \frac{u^2}{[1 - z\bar z]^2}\,.
\end{equation}

Possible further extensions of our work would be the consideration of
holonomies, as well as the inclusion of the second boundary. In particular,
it would be interesting to verify the argumentation
of \cite{Strominger:2001pn},
that the holographic dual is a field theory on one boundary, rather than two.
This is currently under investigation.

\section*{Acknowledgements}
\small

This work was partially supported by INFN, MURST and
by the European Commission RTN program
HPRN-CT-2000-00131, in which the authors are associated to the University of
Torino.
\normalsize

\begin{appendix}

\section{Conventions}

Our SL$(2,\bC)$ generators are

\begin{equation}
\tau_0 = \frac 12\left(\begin{array}{cc} -i & 0 \\ 0 & i \end{array}\right)\,,
         \qquad
\tau_1 = \frac 12\left(\begin{array}{cc} 0 & 1 \\ 1 & 0 \end{array}\right)\,,
         \qquad
\tau_2 = \frac 12\left(\begin{array}{cc} 0 & -i \\ i & 0 \end{array}\right)\,.
\end{equation}

They satisfy

\begin{equation}
[\tau_a, \tau_b] = \epsilon_{abc}\tau^c\,,
\end{equation}

with $\epsilon_{012} = +1$, and are normalized according to

\begin{equation}
{\mathrm{Tr}}(\tau_a\tau_b) = \frac 12 \eta_{ab}\,,
\end{equation}

where $(\eta_{ab}) = {\mathrm{diag}}(-1,1,1)$.\\
Another useful property is

\begin{equation}
\tau_a^{\dag} = \sigma \tau_a \sigma\,,
\end{equation}

with $\sigma \in$ SL$(2,\bC)$ given by

\begin{equation}
\sigma = \left(\begin{array}{cc} i & 0 \\ 0 & -i \end{array}\right)\,.
         \label{sigma}
\end{equation}

We further define

\begin{equation}
\tau_{\pm} = \frac 12 (\tau_1 \mp i\tau_2)\,.
\end{equation}

Finally, $d\tau \wedge d\phi \wedge dy$ is chosen to have positive
orientation.

\end{appendix}


\newpage

\begin{thebibliography}{99}

\bibitem{Hull:1998vg}
C.~M.~Hull,
``Timelike T-duality, de~Sitter space, large N gauge theories and
topological field theory,''
JHEP {\bf 9807} (1998) 021
[hep-th/9806146].

\bibitem{Hull:2000mt}
C.~M.~Hull and R.~R.~Khuri,
``Worldvolume theories, holography, duality and time,''
Nucl.\ Phys.\ B {\bf 575} (2000) 231
[hep-th/9911082].

\bibitem{Banks:2000fe}
T.~Banks,
``Cosmological breaking of supersymmetry,''
[hep-th/0007146];
T.~Banks and W.~Fischler,
``M-theory observables for cosmological space-times,''
[hep-th/0102077].

\bibitem{Balasubramanian:2001rb}
V.~Balasubramanian, P.~Horava and D.~Minic,
``Deconstructing de~Sitter,''
JHEP {\bf 0105} (2001) 043
[hep-th/0103171].

\bibitem{Witten:2001kn}
E.~Witten,
``Quantum gravity in de~Sitter space,''
[hep-th/0106109].

\bibitem{Bousso:2000nf}
R.~Bousso,
``Positive vacuum energy and the N-bound,''
JHEP {\bf 0011} (2000) 038
[hep-th/0010252];
``Bekenstein bounds in de Sitter and flat space,''
JHEP {\bf 0104} (2001) 035
[hep-th/0012052].

\bibitem{Volovich:2001rt}
A.~Volovich,
``Discreteness in de~Sitter space and quantization of Kaehler manifolds,''
[hep-th/0101176].

\bibitem{Chamblin:2001dx}
A.~Chamblin and N.~D.~Lambert,
``De~Sitter space from M-theory,''
Phys.\ Lett.\ B {\bf 508} (2001) 369
[hep-th/0102159];
``Zero-branes, quantum mechanics and the cosmological constant,''
[hep-th/0107031].

\bibitem{Strominger:2001pn}
A.~Strominger,
``The dS/CFT correspondence,''
[hep-th/0106113].

\bibitem{Klemm:2001ea}
D.~Klemm,
``Some aspects of the de Sitter/CFT correspondence,''
[hep-th/0106247].

\bibitem{Mazur:2001aa}
P.~O.~Mazur and E.~Mottola,
``Weyl cohomology and the effective action for conformal anomalies,''
[hep-th/0106151];
I.~Antoniadis, P.~O.~Mazur and E.~Mottola,
``Comment on {\it Nongaussian isocurvature perturbations from inflation},''
[astro-ph/9705200].

\bibitem{Li:2001ky}
M.~Li,
``Matrix model for de~Sitter,''
[hep-th/0106184].

\bibitem{Nojiri:2001mf}
S.~Nojiri and S.~D.~Odintsov,
``Conformal anomaly from dS/CFT correspondence,''
[hep-th/0106191];
``Quantum cosmology, inflationary brane-world creation and dS/CFT
correspondence,''
[hep-th/0107134];
S.~Nojiri, S.~D.~Odintsov and S.~Ogushi,
``Cosmological and black hole brane world universes in higher derivative
gravity,''
[hep-th/0108172].

\bibitem{Gao:2001sr}
Y.~Gao,
``Symmetries, matrices, and de Sitter gravity,''
[hep-th/0107067].

\bibitem{Shiromizu:2001bg}
T.~Shiromizu, D.~Ida and T.~Torii,
``Gravitational energy, dS/CFT correspondence and cosmic no-hair,''
[hep-th/0109057].

\bibitem{Kallosh:2001tm}
R.~Kallosh,
``N = 2 supersymmetry and de Sitter space,''
[hep-th/0109168].

\bibitem{Hull:2001ii}
C.~M.~Hull,
``De Sitter Space in Supergravity and M Theory,''
[hep-th/0109213].

\bibitem{Spradlin:2001pw}
M.~Spradlin, A.~Strominger and A.~Volovich,
``Les Houches Lectures on De Sitter Space,''
[hep-th/0110007].

\bibitem{Perlmutter:2000rr}
S.~Perlmutter,
``Supernovae, dark energy, and the accelerating universe: The status of
the cosmological parameters,''
in {\it Proc. of the 19th Intl. Symp. on Photon and Lepton Interactions
at High Energy LP99 } ed. J.A. Jaros and M.E. Peskin,
Int.\ J.\ Mod.\ Phys.\ A {\bf 15S1} (2000) 715
[eConfC {\bf 990809} (2000) 715].

\bibitem{Maldacena:1998ih}
J.~Maldacena and A.~Strominger,
``Statistical entropy of de~Sitter space,''
JHEP {\bf 9802} (1998) 014
[gr-qc/9801096].

\bibitem{Park:1998qk}
M.~I.~Park,
``Statistical entropy of three-dimensional Kerr-de Sitter space,''
Phys.\ Lett.\ B {\bf 440} (1998) 275
[hep-th/9806119].

\bibitem{Banados:1999tb}
M.~Ba\~{n}ados, T.~Brotz and M.~E.~Ortiz,
``Quantum three-dimensional de~Sitter space,''
Phys.\ Rev.\ D {\bf 59} (1999) 046002
[hep-th/9807216].

\bibitem{Kim:1999zs}
W.~T.~Kim,
``Entropy of 2+1 dimensional de Sitter space in terms of brick wall method,''
Phys.\ Rev.\ D {\bf 59} (1999) 047503
[hep-th/9810169].

\bibitem{Lin:1999gf}
F.~Lin and Y.~Wu,
``Near-horizon Virasoro symmetry and the entropy of de Sitter space in
any dimension,''
Phys.\ Lett.\ B {\bf 453} (1999) 222
[hep-th/9901147].

\bibitem{Hawking:2001da}
S.~W.~Hawking, J.~Maldacena and A.~Strominger,
``DeSitter entropy, quantum entanglement and AdS/CFT,''
JHEP {\bf 0105} (2001) 001
[hep-th/0002145].

\bibitem{'tHooft:1993gx}
G.~'t Hooft,
``Dimensional reduction in quantum gravity,''
[gr-qc/9310026].

\bibitem{Susskind:1995vu}
L.~Susskind,
``The World as a hologram,''
J.\ Math.\ Phys.\  {\bf 36} (1995) 6377
[hep-th/9409089].

\bibitem{Maldacena:1998re}
J.~Maldacena,
``The large $N$ limit of superconformal field theories and supergravity,''
Adv.\ Theor.\ Math.\ Phys.\  {\bf 2} (1998) 231
[Int.\ J.\ Theor.\ Phys.\  {\bf 38} (1998) 1113]
[hep-th/9711200].

\bibitem{Bousso:1999cb}
R.~Bousso,
``Holography in general space-times,''
JHEP {\bf 9906} (1999) 028
[hep-th/9906022].

\bibitem{Brown:1986nw}
J.~D.~Brown and M.~Henneaux,
``Central charges in the canonical realization of asymptotic symmetries:
An example from three-dimensional gravity,''
Commun.\ Math.\ Phys.\  {\bf 104} (1986) 207.

\bibitem{Deser:dr}
S.~Deser and R.~Jackiw,
``Three-Dimensional Cosmological Gravity: Dynamics Of Constant Curvature,''
Annals Phys.\  {\bf 153} (1984) 405.

\bibitem{Mazur:qc}
P.~O.~Mazur,
``Quantum Gravitational Measure For Three Geometries,''
Phys.\ Lett.\ B {\bf 262} (1991) 405
[hep-th/9701033].

\bibitem{Achucarro:1986vz}
A.~Achucarro and P.~K.~Townsend,
``A Chern-Simons Action For Three-Dimensional Anti-De Sitter
Supergravity Theories,''
Phys.\ Lett.\ B {\bf 180} (1986) 89.

\bibitem{Witten:1988hc}
E.~Witten,
``(2+1)-Dimensional Gravity As An Exactly Soluble System,''
Nucl.\ Phys.\ B {\bf 311} (1988) 46.

\bibitem{Coussaert:1995zp}
O.~Coussaert, M.~Henneaux and P.~van Driel,
``The Asymptotic dynamics of three-dimensional Einstein gravity
with a negative cosmological constant,''
Class.\ Quant.\ Grav.\  {\bf 12} (1995) 2961
[gr-qc/9506019].

\bibitem{Banados:1998pi}
M.~Ba\~{n}ados, K.~Bautier, O.~Coussaert, M.~Henneaux and M.~Ortiz,
``Anti-de Sitter/CFT correspondence in three-dimensional supergravity,''
Phys.\ Rev.\ D {\bf 58} (1998) 085020
[hep-th/9805165].

\bibitem{Henneaux:2000ib}
M.~Henneaux, L.~Maoz and A.~Schwimmer,
``Asymptotic dynamics and asymptotic symmetries of three-dimensional
extended AdS supergravity,''
Annals Phys.\  {\bf 282} (2000) 31
[hep-th/9910013].

\bibitem{Rooman:2001zi}
M.~Rooman and P.~Spindel,
``Holonomies, anomalies and the Fefferman-Graham ambiguity in AdS(3)
gravity,''
Nucl.\ Phys.\ B {\bf 594} (2001) 329
[hep-th/0008147].

\bibitem{Krasnov:2001cu}
K.~Krasnov,
``On holomorphic factorization in asymptotically AdS 3D gravity,''
[hep-th/0109198].

\bibitem{Moore:1989yh}
G.~W.~Moore and N.~Seiberg,
``Taming The Conformal Zoo,''
Phys.\ Lett.\ B {\bf 220} (1989) 422;
S.~Elitzur, G.~W.~Moore, A.~Schwimmer and N.~Seiberg,
``Remarks On The Canonical Quantization Of The Chern-Simons-Witten Theory,''
Nucl.\ Phys.\ B {\bf 326} (1989) 108.

\bibitem{Alekseev:1989ce}
A.~Alekseev and S.~Shatashvili,
``Path Integral Quantization Of The Coadjoint Orbits Of The Virasoro
Group And 2-D Gravity,''
Nucl.\ Phys.\ B {\bf 323} (1989) 719.

\bibitem{Bershadsky:1989mf}
M.~Bershadsky and H.~Ooguri,
``Hidden Sl(N) Symmetry In Conformal Field Theories,''
Commun.\ Math.\ Phys.\  {\bf 126} (1989) 49.

\bibitem{Forgacs:1989ac}
P.~Forgacs, A.~Wipf, J.~Balog, L.~Feher and L.~O'Raifeartaigh,
``Liouville And Toda Theories As Conformally Reduced WZNW Theories,''
Phys.\ Lett.\ B {\bf 227} (1989) 214.

\bibitem{Sabra:1993hw}
W.~A.~Sabra,
``Classical Hamiltonian reduction and superconformal algebras,''
Phys.\ Lett.\ B {\bf 313} (1993) 68
[hep-th/9207012].

\bibitem{Seiberg:1990eb}
N.~Seiberg,
``Notes On Quantum Liouville Theory And Quantum Gravity,''
Prog.\ Theor.\ Phys.\ Suppl.\  {\bf 102} (1990) 319.

\end{thebibliography}
\end{document}